\documentclass[twocolumn,showpacs,preprintnumbers,amsmath,amssymb]{revtex4}

\usepackage{graphicx}
\usepackage{dcolumn}
\usepackage{bm}

\begin{document}


\title{Quantum phases of dipolar spinor condensates}
\author{S. Yi$^1$, L. You$^2$, and H. Pu$^1$}

\affiliation{$^1$Department of Physics and Astronomy, and Rice
Quantum Institute, Rice University, Houston, TX 77251-1892, USA \\
$^2$School of Physics, Georgia Institute of Technology, Atlanta,
GA 30332-0430, USA}

\begin{abstract}
We study the zero-temperature ground state structure of a spin-1
condensate with magnetic dipole-dipole interactions. We show that
the dipolar interactions break the rotational symmetry of the
Hamiltonian and induce new quantum phases. Different phases can be
reached by tuning the effective strength of the dipolar
interactions via modifying the trapping geometry. The experimental
feasibility of detecting these phases is investigated. The
spin-mixing dynamics is also studied.
\end{abstract}

\date{\today}

\maketitle

A dipolar condensate is a condensate possessing the dipole-dipole
interactions, in addition to the usual $s$-wave contact
interaction. Recent theoretical studies have found that, the
long-range dipolar interactions (which may originate either from
the intrinsic atomic \cite{dipoleatom} or molecular
\cite{damski,bethlem,weinstein} magnetic dipole moments or from
the induced electric dipole moments \cite{yi1}) may greatly affect
the stability of the condensate \cite{yi1,goral1,santos,giov} and
significantly modify the excitation spectrum
\cite{yi2,goral2,dell,santos2}. Furthermore, the dipolar
interactions are responsible for the generation of translational
entanglement \cite{opatrny} and a variety of novel quantum phases
\cite{goral3} in condensates confined in optical lattices. All
these works have focused on scalar condensates. The effect of the
dipolar interactions on spinor condensates confined in multi-well
potentials is explored in the works of Refs.~\cite{pumag}, where
only the the dipolar interactions between different potential
wells are taken into account. In this Letter, we study the
properties of dipolar spinor condensates in a single trap. We show
that the interplay between the spin-exchange and the dipolar
interactions gives rise to extremely rich physics unexplored so
far and that as the effective dipolar strength depends on the
condensate aspect ratio, transitions between new quantum phases
resulting from the dipolar interactions can be induced by
modifying the trap geometry.

The intrinsic magnetic dipole moment of an atom is related to its
total angular momentum as ${\boldsymbol\mu}_F=-g_F\mu_B{\mathbf F}
$, with $g_F$ being the Land\'{e} $g$-factor, $\mu_B$ the Bohr
magneton, and ${\mathbf F}$ the angular momentum operator. Thus
the dipole-dipole interactions will affect the condensate once the
spin degrees of freedom become accessible as in an optical trap.
Since the first creation of spinor condensates in dilute alkali
atomic vapors \cite{kurn}, we have witnessed tremendous
experimental and theoretical studies on these systems
\cite{barret,chang,gorlitz,ho,law,burke,goldstein,pu2}. The key
feature of the spinor condensate is the spin-exchange interaction
which is responsible for a variety of interesting physics such as
the spin-mixing \cite{law}, phase conjugation \cite{goldstein},
spin domain formation \cite{kurn} and topological defects
generation \cite{pu2}. All these studies, however, neglected the
dipolar interactions, except in Ref. \cite{gu}, where Gu suggested
that the internal dipolar field may induce spontaneous
magnetization in a spinor condensate. Nevertheless, the dipolar
interactions are not treated explicitly in that work.

Consider a spin-1 condensate with $N$ atoms. The Hamiltonian
without the dipolar intereactions, in second quantized form, reads
\cite{ho}
\begin{eqnarray*}
H_{\rm sp}&=&\int d{\mathbf r}\hat\psi_\alpha^\dag({\mathbf
r})\left(-\frac{\hbar^2\nabla^2}{2M}+V_{\rm
ext}\right)\hat\psi_\alpha({\mathbf r}) \\
&&+\frac{c_0}{2}\int d{\mathbf r}\hat\psi_\alpha^\dag({\mathbf
r})\hat\psi_\beta^\dag({\mathbf r}) \hat\psi_\alpha({\mathbf
r})\hat\psi_\beta({\mathbf r}) \\
&&+\frac{c_2}{2}\int d{\mathbf r}\hat\psi_\alpha^\dag({\mathbf r})
\hat\psi_{\alpha'}^\dag({\mathbf r}){\mathbf F}_{\alpha\beta}\cdot
{\mathbf F}_{\alpha'\beta'}\hat\psi_\beta({\mathbf
r})\hat\psi_{\beta'}({\mathbf r}),
\end{eqnarray*}
where $M$ is the mass of the atom, $\hat\psi_\alpha({\mathbf r})$
($\alpha=0,\pm1$) denotes the annihilation operator for the
$m_F=\alpha$ component of a spin-1 field. The trapping potential
$V_{\rm ext}$ is assumed to be spin-independent. The collisional
interaction parameters are $c_0=4\pi\hbar^2(a_0+2a_2)/(3M)$ and
$c_2=4\pi\hbar^2(a_2-a_0)/(3M)$ with $a_f$ ($f=0,2$) being the
$s$-wave scattering length for spin-1 atoms in the combined
symmetric channel of total spin $f$. For the two experimentally
realized spinor condensate systems ($^{23}$Na and $^{87}$Rb), we
have $|c_2| \ll c_0$.

The Hamiltonian for the dipolar interactions reads
\begin{eqnarray*}
H_{\rm dd}&=&\frac{c_d}{2}\int d{\mathbf r} \int d{\mathbf
r}'\,\frac{1}{|{\mathbf r}-{\mathbf
r}'|^3} \\
&&\times\left[\hat\psi_\alpha^\dag({\mathbf
r})\hat\psi_{\alpha'}^\dag({\mathbf r}'){\mathbf
F}_{\alpha\beta}\cdot{\mathbf
F}_{\alpha'\beta'}\hat\psi_\beta({\mathbf
r})\hat\psi_{\beta'}({\mathbf
r}')\right. \\
&&\left.-3\hat\psi_\alpha^\dag({\mathbf
r})\hat\psi_{\alpha'}^\dag({\mathbf r'}){\mathbf
F}_{\alpha\beta}\cdot {\mathbf e}\,{\mathbf
F}_{\alpha'\beta'}\cdot{\mathbf e}\,\hat\psi_\beta({\mathbf
r})\hat\psi_{\beta'}({\mathbf r}')\right],
\end{eqnarray*}
where $c_d=\mu_0\mu_B^2g_F^2/(4\pi)$ and ${\mathbf e}=({\mathbf
r}-{\mathbf r}')/|{\mathbf r}-{\mathbf r}'|$ is a unit vector. The
total Hamiltonian is then $ H_{\rm tot}=H_{\rm sp}+H_{\rm dd} $,
from which we can derive the Hesenberg equations for the boson
field operators. In the standard mean-field treatment, one
replaces the field operators $\hat\psi_\alpha$ by their
expectation values
$\phi_\alpha\equiv\langle\hat\psi_\alpha\rangle$, which yields the
so-called Gross-Pitaevskii equations (GPEs).

Before proceeding further, let us first estimate the relative
strengths of the spin-exchange and the dipolar interactions which
are characterized by $c_2$ and $c_d$, respectively. If we take
$a_0=50 a_B$ ($a_B$ being the Bohr radius) and $a_2=55 a_B$ for
$^{23}$Na \cite{burke}, we find $c_d/|c_2|\simeq 0.007$.
Similarly, for $^{87}$Rb ($a_0=101.8 a_B$ and $a_2=100.4a_B$
\cite{burke}), we have $c_d/|c_2|\simeq 0.1$. Hence at least for
$^{87}$Rb, the dipolar interaction is not negligible compared to
the spin-exchange interaction.

The facts that $|c_2|\ll c_0$ and $c_d\ll |c_2|$ inspire us to
invoke the single mode approximation (SMA) \cite{law}, namely, $
\hat\psi_\alpha({\mathbf r})\approx\phi({\mathbf r})\hat
a_\alpha$, where $\phi({\mathbf r})$ is the {\em spin-independent}
condensate spatial wave function, $\hat a_\alpha$ is the
annihilation operator for $m_F=\alpha$ component. The validity of
the SMA can be checked by solving the GPEs numerically for the
ground state wave functions $\phi_\alpha$. We have performed these
calculations and found that the SMA is valid for $c_d \lesssim
0.2|c_2|$ with scattering lengths of $^{23}$Na and $^{87}$Rb.

The Hamiltonian $H_{\rm sp}$ under the SMA is given by (after
dropping spin-independent constant terms) \cite{law}:
\begin{eqnarray}
H_{\rm sp} = c_2'\, \hat{\mathbf L}^2, \label{hsp1}
\end{eqnarray}
where $c_{2}'=(c_{2}/2)\int d{\mathbf r}|\phi({\mathbf r})|^4$ is
the spin-exchange interaction coefficient, and $\hat{\mathbf
L}=\hat a_\alpha^\dag{\mathbf F}_{\alpha\beta}\hat a_\beta$ is the
total many-body angular momentum operator of the system, whose
eigenstates and eigenvalues are defined by
\begin{eqnarray*}
\hat{\mathbf L}^2|l,m\rangle=l(l+1)|l,m\rangle,\;\;\; \hat
L_z|l,m\rangle=m|l,m\rangle,
\end{eqnarray*}
where $m=0,\pm1,\ldots,\pm l$ and for a given total number of
atoms $N$, the allowable values of $l$ are $l=0,2,4,\ldots,N$ for
even $N$ and $l=1,3,5,\ldots, N$ for odd $N$. The ground state
properties of the spinor condensate under $H_{\rm sp}$ is
therefore completely determined by the sign of $c_2'$ \cite{law}:
For $c_2'> 0$ (antiferromagnetic), the ground state is given by
the spin singlet $|G \rangle = |0,0\rangle$; for $c_2' < 0$
(ferromagnetic), it is given by $|G\rangle = |N,m\rangle$ which
has a $(2N+1)$-fold degeneracy with $m$ taking any integer numbers
between $-N$ and $N$.

The dipolar interaction Hamiltonian under the SMA is given by
\begin{eqnarray}
H_{\rm dd}&=&\frac{c_d}{2}\int d{\mathbf r}\int d{\mathbf
r}'\frac{|\phi({\mathbf r})|^2|\phi({\mathbf r}')|^2}{|{\bf
r}-{\bf r}'|^3}\Big[\left(\hat{\bf L}^2-3(\hat{\mathbf
L}\cdot{\mathbf
e})^2\right)\nonumber\\
&& -\left(2N-3\hat a_\alpha^\dag{\mathbf F}_{\alpha\beta}
\cdot{\mathbf e}\,{\mathbf F}_{\beta\beta'}\cdot{\mathbf e}\,\hat
a_{\beta'}\right)\Big].\label{hdipq1}
\end{eqnarray}
In general, this is still in a very complicated form. Remarkably,
for a condensate with axial symmetry (which happens to be the most
experimentally relevant case), with its symmetry axis chosen to be
along the quantization axis, $z$, $H_{\rm dd}$ takes a very simple
form:
\begin{equation}
H_{\rm dd} = -c_d' \hat{\bf L}^2 +3c_d' (\hat L_z^2 + \hat n_0),
\label{hddsma}
\end{equation}
where $\hat n_0= \hat a_0^\dagger \hat a_0$ is the number operator
for $m_F=0$ and $c_d'=(c_d/4)\int \int d{\mathbf r}d{\mathbf
r}'|\phi({\mathbf r})|^2|\phi({\mathbf
r}')|^2(1-3\cos^2\theta_{\mathbf e})/|{\mathbf r}-{\mathbf r}'|^3$
with $\theta_{\mathbf e}$ being the polar angle of $({\mathbf
r}-{\mathbf r}')$. The total Hamiltonian under the SMA is then
\begin{equation}
H_{\rm tot} = (c_2'-c_d') \hat {\bf L}^2 +3c_d' (\hat L_z^2 + \hat
n_0). \label{qmham}
\end{equation}

Before we discuss the ground state of $H_{\rm tot}$, we want to
point out that $H_{\rm sp}$ in (\ref{hsp1}) possesses SO(3)
symmetry, i.e., it is rotationally invariant in spin space
\cite{ho}. The presence of the dipolar interaction, however,
breaks this symmetry. This is not unlike the situation in
superfluid $^3$He, where the dipolar interaction between nuclear
spins, despite of its smallness, breaks the spin-orbit symmetry
and is crucial for the understanding of the superfluid phases of
the system \cite{3he}. As we shall see below, the dipolar
interaction of atomic spins also results in new quantum phases in
spinor condensate.

From Hamiltonian (\ref{qmham}), one can expect that the behavior
of the dipolar spinor condensate should be very sensitive to the
signs of both $(c_2'-c_d')$ and $c_d'$. Unlike $c_2'$, both the
sign and the magnitude of the dipolar interaction coefficient
$c_d'$ depend on the geometric shape of the condensate, which
makes the system highly tunable through the modification of the
trap aspect ratio. As an example, we assume the condensate wave
function has a Gaussian form $\phi({\mathbf
r})=\pi^{-3/4}\kappa^{1/2}
q^{-3/2}e^{-(x^2+y^2+\kappa^2z^2)/(2q^2)}$, simple calculation
shows that \cite{yi1},
\begin{eqnarray}
c\equiv c_d'/|c_2'|=2\pi c_d\,\chi(\kappa)/(3|c_2|),
\end{eqnarray}
where $\chi(\kappa)$ is a monotonically increasing function of the
condensate aspect ratio $\kappa$, bounded between $-1$ and $2$,
and passing through zero at $\kappa=1$ \cite{gaussian}. From this
it is clear that both the sign and the strength of the effective
dipolar interaction can be tuned with trapping geometry.

\begin{figure}[h]
\centering
\includegraphics[width=2.6in]{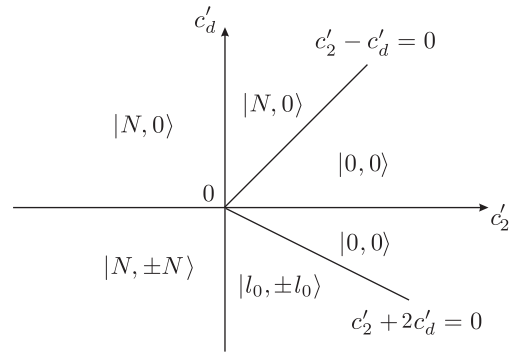}
\caption{The ground states of $H_0$, where $l_0=N$ if
$N>(c-1)/(2c+1)$, otherwise $l_0=0$. For typical parameters, the
inequality is usually satisfied, so we can always take $l_0=N$.}
\label{domain}
\end{figure}

To gain more insights into the structure of the ground state, let
us first neglect the $n_0$-term in $H_{\rm tot}$. The remaining
Hamiltonian, denoted by $H_0$ [$=(c_2'-c_d') \hat{\bf L}^2 +3c_d'
\hat L_z^2 $], has a diagonalized form under the basis states
$|l,m\rangle$. The ground state of $H_0$ can be easily found and
the phase diagram is illustrated in Fig.~\ref{domain}. We see that
the ground state of $H_0$ is divided into three regimes in the
$c_2'$-$c_d'$ parameter space, and the ground state angular
momentum quantum number $l$ can be either $N$ or $0$.

For the case of $l=N$, we expect the ground state of the total
Hamiltonian $H_{\rm tot}$ to be very close to that of $H_0$, since
the $n_0$-term is at least a factor of $1/N$ smaller than the
rest. In contrast, for the $l=0$ case, the $n_0$-term is expected
to be important and the true ground state should deviate from
$|0,0\rangle$ significantly unless $c=0$. We found the ground
state of $H_{\rm tot}$ numerically by expanding the Hamiltonian
onto a Fock state basis $|N_{-1}, N_0, N_1\rangle$ with the
constraint $N=\sum_\alpha N_\alpha$, and the calculations
confirmed our qualitative arguments above. In the following, we
will discuss the properties of the ground state in more detail and
we will present our results for the cases $c_2' > 0$ and $c_2' <0$
separately.

\begin{figure}[h]
\centering
\includegraphics[width=2.6in]{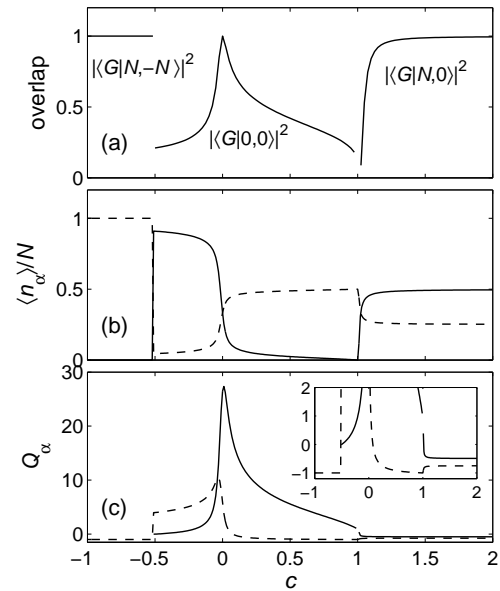}
\caption{The $c$-dependence of the overlap between the true ground
state $|G\rangle$ and that of $H_0$ (a), the normalized ground
state population (b), and the number fluctuation (c). The solid
and dashed lines in (b) and (c) denote the $m_F=0$ and $-1$
components respectively. The inset in (c) shows the detailed
behavior of $Q_\alpha$ at the vicinity of $Q_\alpha=0$. The
results presented here are for $c_2'>0$ and $N=100$
\cite{numerics}.} \label{gndc2gt0}
\end{figure}

$c_2' > 0$ {\em case}: Figure \ref{gndc2gt0} (a) illustrates the
overlap between the true ground state (denoted by $|G\rangle$) and
that of $H_0$. The overlap is unit in the regions $c<-1/2$ and it
quickly approaches unit for $c>1$. For $-1/2 < c < 1$, the overlap
decreases rapidly from 1 when $c$ deviates from 0. In this region,
in general $|G\rangle = \sum_l g_l |l,0\rangle$, with the
coefficients $g_l$ dependent upon $c$, is a superposition of
different angular momentum states with $\langle \hat{L}_z \rangle
=0$. The normalized ground state population and its variance are
presented in Fig.~\ref{gndc2gt0} (b) and (c). Two sharp phase
boundaries can be seen at $c=-1/2$ and $c=1$ in agreement with
Fig.~\ref{domain}. The variance is presented in terms of Mandel
$Q$-factor. $Q<0$, $=0$, and $>0$ represent sub-Poissonian,
Poissonian, and super-Poissonian distributions, respectively. In
particular, $Q=-1$ represents a Fock state with vanishing number
fluctuations.

For $c<-1/2$, $|G\rangle =|N,\pm N\rangle$ is a Fock state with
all the population in either $m_F=1$ or $-1$ state. This is the
state with spontaneous magnetization discussed in Ref.~\cite{gu}.
Since the ground state in this regime is two-fold degenerated with
opposite magnetization, a chain of such condensates will form an
example of the Ising model of statistical mechanics. For $c>1$,
$|G\rangle \approx |N,0\rangle$. Expanded onto the Fock state
basis $|N_{-1},N_0,N_1\rangle$, we have $|N,0\rangle
=\sum_{k=0}^{N/2} g_k |N/2-k, 2k, N/2-k\rangle$ (assuming $N$ to
be even), where for $N \gg 1$, the expansion coefficients $g_k =
(8/N\pi^2)^{1/4}\exp [- 4(k-N/4)^2/N]$, from which we immediately
have $\langle \hat{n}_0 \rangle =2\langle \hat{n}_{\pm1} \rangle
=N/2$ , $Q_0=-1/2$ and $Q_{\pm 1} =-3/4$. Hence the populations in
all three spin components have sub-Poissonian distributions.
Finally, for $-1/2 <c<1$, population in $m_F=0$ state has a
super-Poissonian fluctuation with maximum fluctuations occur at
$c=0$, while the population fluctuations in $m_F= \pm 1$ approach
the Fock state limit for positive $c$.

$c_2' < 0$ {\em case}: For this case, the overlap between the true
ground state and that of $H_0$ is nearly perfect. The phase
diagram is characterized by a sharp boundary at $c=0$, and the
ground state is given by the maximally polarized state $|N, \pm N
\rangle$ or $|N,0\rangle$ for $c<0$ and $c>0$, respectively. Since
the phase boundary occurs at $c=0$ (vanishing dipolar
interactions), this phase transition should be readily verifiable
in $^{87}$Rb spinor condensates by changing the condensate shape
from cigar to pancake, or vice versa. As we have mentioned
earlier, without the dipolar interaction, the ground state of a
ferromagnetic spin-1 condensate is given by $|N,m\rangle$ for any
integer $m$ between $-N$ and $N$ (i.e., the spin vector has no
preferred spatial orientation). Now the dipolar interaction breaks
this degeneracy and orients the spin vector along (perpendicular
to) the axial direction for a cigar-shaped (pancake-shaped)
condensate, a clear manifestation of the symmetry-breaking
properties of the dipole force.

\begin{figure}[h]
\centering
\includegraphics[width=2.6in]{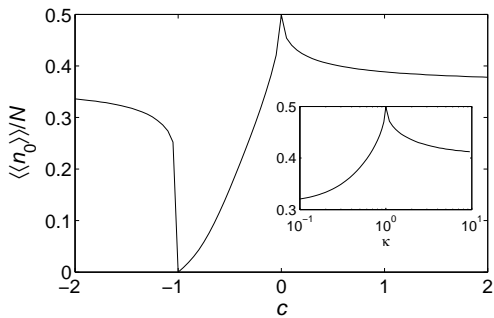}
\caption{The $c$-dependence of the steady state population of
$m_F=0$ component for the initial state $|N/2,0,N/2\rangle$ with
$N=10^4$ and $c_2'<0$. The inset indicates its $\kappa$-dependence
for $^{87}$Rb condensate with a Gaussian wave function.}
\label{timeave}
\end{figure}

In addition to the ground state phase structures, the spin-mixing
dynamics \cite{law} of the dipolar spinor condensate can also be
studied by numerically evolving an initial state under $H_{\rm
tot}$. Figure~\ref{timeave} illustrates one example. Here the
initial state is given by the Fock state $|N/2,0,N/2\rangle$ with
half population in $m_F=1$ and $-1$ components, respectively. The
spin-mixing dynamics will quickly drive the system into a
quasi-steady state. Again we want to focus on the effect of the
dipolar interactions whose strength depends on the condensate
aspect ratio $\kappa$. In Fig.~\ref{timeave} the steady state
population of spin-0 is plotted as a function of $c$, while the
inset replots the population as a function of $\kappa$ using
parameters for $^{87}$Rb. As we can see, modifying the condensate
aspect ratio changes significantly the population distribution in
the steady state.

In summary, we have studied the ground state properties and
spin-mixing dynamics of a dipolar spinor condensate. We have shown
that the dipolar interaction breaks the rotational symmetry of the
Hamiltonian and as a result the ground state is characterized by
several distinct quantum phases depending on the relative
strengths of the spin-exchange and dipolar interactions. The
transitions between these phases may be induced by simply
modifying the trapping geometry of the condensate.

We have neglected any external magnetic field in our study here.
The presence of external fields will affect the orientation of the
spin and hence change the phase diagram. For the change to be
insignificant, we need to reduce the field strength such that the
zeeman energy is weaker than the dipolar energy. For typical
values of alkali atoms, this requires to control the magnetic
field below $10^{-4}$ Gauss \cite{bfield}. This requirement will
pose an experimental challenge (none of the experiments on spinor
BEC so far \cite{kurn,barret,chang,gorlitz} has met this
field-free requirement), but is within reach with current
technologies. We hope our work will stimulate experimental efforts
along this line. The effect of the magnetic field on dipolar
spinor condensates is currently under study. Our future works will
also include the study of spin-2 dipolar spinor condensate
\cite{chang,gorlitz}. We believe that these studies will open many
unexplored and promising avenues of research in the field of
quantum degenerate atomic gases.

We thank Jason Ho and Michael Chapman for helpful discussions.


\begin{thebibliography}{}
\bibitem{numerics} Obviously the maximum total number of atoms used in
the calculations are limited by computer power. However, we have
found essentially the same behavior for $N$ from a few tens to
around 100. So we believe the results are valid for realistic
atomic numbers ($10^4 \sim 10^6$).
\bibitem{bfield} This requirement can be relaxed a little if atoms
with larger magnetic moment (e.g., Cr, Sr, etc.) can be used. But
none of these atoms has been condensed so far.
\end{thebibliography}
\end{document}